\documentstyle[preprint,eqsecnum,aps]{revtex}
\tightenlines
\newcommand{\be}{\begin{equation}}
\newcommand{\ee}{\end{equation}}
\newcommand{\ba}{\begin{eqnarray}}
\newcommand{\ea}{\end{eqnarray}}

\begin{document}
\draft

\title{Statistical Entropy of a Stationary Dilaton Black Hole\\ from
 Cardy Formula}
 \author{Jiliang Jing $^{*\ a\ b }$\footnotetext[1]
 {email: jljing@hunnu.edu.cn} \ \
 \ \ Mu-Lin Yan $^{\dag\ b}$\footnotetext[2]
 {email: mlyan@ustc.edu.cn}}
\address{a) Physics Department and Institute of Physics , Hunan Normal
University,\\ Changsha, Hunan 410081, P. R. China;  \\ b)
Department of Astronomy and Applied Physics, University of Science
and Technology of China, \\ Hefei, Anhui 230026, P. R. China}

\maketitle
\begin{abstract}
With Carlip's boundary conditions, a standard Virasoro subalgebra
with corresponding central charge for stationary dilaton black
hole obtained in the low-energy effective field theory describing
string is constructed at a Killing horizon. The statistical
entropy of stationary dilaton black hole yielded by standard Cardy
formula agree with its Bekenstein-Hawking entropy only if we take
period $ T$ of function $v$ as the periodicity of the Euclidean
black hole. On the other hand, if we consider first-order quantum
correction then the entropy contains a logarithmic term with a
factor $-\frac{1}{2}$, which is different from Kaul and Majumdar's
one, $-\frac{3}{2}$. We also show that the discrepancy is not just
for the dilaton black hole, but for any one whose corresponding
central change  takes the form $\frac{c}{12}= \frac{A_H}{8\pi
G}\frac{2\pi}{\kappa T}$.
 \vspace*{0.5cm}
 \end{abstract}


\section{INTRODUCTION}
\label{sec:intro} \vspace*{0.0cm}

Much effort has been concentrated on the statistical mechanical
description of the Bekenstein-Hawking black hole entropy
\cite{Bekenstein72}- \cite{Hawking74} in terms of microscope
states both in string theory \cite{Youm99} and in ``quantum
geometry" \cite{Ashtekar98}. Strominger \cite{Strominger98}
calculated the entropy of black holes whose near-horizon geometry
is locally $AdS_3$ from the asymptotic growth of states. Carlip
\cite{Carlip99l} \cite{Carlip99}  derived the central extension of
the constraint algebra of general relativity by using
Brown-Henneaux-Strominger's approach \cite{Strominger98} and
manifestly covariant phase space methods \cite{Lee90}-
\cite{Iyer95}. He found that a natural set of boundary conditions
on the (local) Killing horizon leads to a Virasoro subalgebra with
a calculable central charge and the standard Cardy formula gives
the Bekenstein-Hawking entropies of some black holes. Those works
show a suggestion that the asymptotic behavior of the density of
states may be determined by the algebra of diffeomorphism at
horizon. Solodukhin \cite{Solodukhin99} obtained same result by a
analysis of the Liouville theory near the horizon obtained from
dimensional reduction of Einstein gravity. Das, Ghosh, and Mitra
\cite{Das00} studied the statistical entropy of a Schwarzschild
black string in five dimensions  by counting the black string
states which from a representation of the near-horizon conformal
symmetry with a central charge. Recently, we \cite{Jing00a}
extended Carlip's investigation \cite{Carlip99} for vacuum case to
a case including a cosmological term and electromagnetic fields
and calculated the statistical entropies of the Kerr-Newman black
hole and the Kerr-Newman-AdS black hole  by using standard Cardy
formula.

On the other hand, the quantum correction to entropy of the black
hole is an interesting topic\cite{Kaul00}-\cite{Jing00}. Recently,
Kaul and Majumdar \cite{Kaul00} computed the lowest order
corrections to the Bekenstein-Hawking entropy in a particular
formulation \cite{Kaul98}of the ``quantum geometry" program of
Ashtekar et al. They showed that the leading corrections is a
logarithmic term, i. e., the entropy is
 \be \label{KM}
 S\sim \frac{A_H}{4}-\frac{3}{2}\ln \frac{A_H}{4} + const.+...,
 \ee
where $A_H$ is the event horizon area. Carlip \cite{Carlip00} also
calculated the quantum corrections to black hole entropy by the
Cardy formula and found that the entropy can be expressed as
 \be\label{Centropy}
S\sim S_0-\frac{3}{2}\ln S_0 +\ln c +const.+...,
 \ee
where $S_0$ is standard Bekenstein-Hawking entropy and $c$ is a
central charge of a Virasoro subalgebra. Carlip pointed out that
if the central charge is the sense of being independent of the
horizon area (Carlip thinks that this can be done by adjust the
periodicity $\beta$ \cite{Carlip00}), then the factor of $-3/2$ in
logarithmic term will always appear.

We all know that four dimensional dilaton charged black hole
obtained in the low-energy effective field theory describing
strings have qualitatively different properties from those that
appear in the ordinary Einstein gravity. Therefore, it is worth to
investigate  whether or not the Carlip's conclusion (the
asymptotic behavior of the density of states may be determined by
the algebra of diffeomorphism at horizon) and Kaul and Majumdar's
result (the leading corrections to the entropy is a logarithm of
the horizon area with a factor $-3/2$) are valid for the static
and stationary dilaton black hole.

We begin in Section II by using the covariant phase techniques to
extend Carlip's investigation \cite{Carlip99} for vacuum case
${\mathbf{L}}_{a_1a_2\cdots a_{n}}= \frac{1}{16\pi
G}\epsilon_{a_1a_2\cdots a_{n}}R $ to a case for gravity coupled
to a Maxwell field and a dilaton, i.e., the Lagrangian n-form is
described by ${\mathbf{L}}_{a_1a_2\cdots a_{n}}=
\epsilon_{a_1a_2\cdots
a_{n}}\left[R-2(\nabla\phi)^2-e^{-2\alpha\phi}F^2\right].$ A
constraint algebra is obtained. In Sec. III, the standard Virasoro
subalgebra with corresponding central charges is constructed for
stationary dilation black hole. The statistical entropy of the
black hole is then calculated by using standard Cardy formula. In
Sec. IV, a new Cardy formula is obtained and then the first-order
quantum correction to the entropy is studied. The last section
devotes to discussion and summary.

\section{Algebra of diffeomorphism on the Killing Horizon}
\vspace*{0.0cm}

Let $\xi ^a$ be any smooth vector fields on a spacetime manifold
${\mathbf{M}}$, i. e., $\xi ^a$ is the infinitesimal generator of
a diffeomorphism, Lee, Wald, and Iyer \cite{Lee90} \cite{Wald93}
\cite{Iyer94} \cite{Iyer95} showed that the Lagrangian
${\mathbf{L}}$, equation of motion n-form ${\mathbf{E}}$,
symplectic potential (n-1)-form ${\mathbf{\Theta}}$, Noether
current (n-1)-form ${\mathbf{J}}$, and Noether charge (n-2)-form
${\mathbf{Q}}$ satisfy following relations
 \ba \delta
 {\mathbf{L}}&=&{\mathbf{E}} \delta\phi+d {\mathbf{\Theta}}, \label{l}\\
{\mathbf{J}}[\xi]&=&{\mathbf{\Theta}} [\phi, {\mathcal{L}}_{\xi}
\phi]-{\mathbf{\xi}} \cdot {\mathbf{L}}, \label{j1}\\
 {\mathbf{J}}&=&d {\mathbf{Q}},\label{dQ} \ea
here and hereafter the ``central dot" denotes the contraction of
the vector field $\xi ^a$ into the first index of the differential
form. Hamilton's equation of motion is given by
\begin{equation}
\delta H[\xi]=\int _C \omega [\phi, \delta \phi,
{\mathcal{L}}_{\xi}\phi]=\int _C[\delta
{\mathbf{J}}[\xi]-d({\mathbf{\xi}} \cdot {\mathbf{\Theta}} [\phi,
\delta \phi])]. \label{dh}
\end{equation}
By using Eq. (\ref{dQ}) and defining a (n-1)-form ${\mathbf{B}}$
as
\be
\delta \int _{\partial C} {\mathbf{\xi}} \cdot
 {\mathbf{ B}}[\phi]=\int _{\partial C}{\mathbf{\xi}} \cdot
 {\mathbf{\Theta}} [\phi. \delta \phi],\label{bq}
\ee
 the Hamiltonian can be expressed as \cite{Carlip99}
\begin{equation}\label{H}
H[\xi]=\int _{\partial C}({\mathbf{Q}}[\xi]-{\mathbf{\xi}}\cdot
{\mathbf{B}}[\phi]).
\end{equation}
The Poisson bracket forms a standard ``surface deformation
algebra" \cite{Brown86} \cite{Carlip99}
\begin{equation}\label{algeb}
  \{H[\xi_1], H[\xi _2]\}=H[\{\xi_1, \xi_2\}]+K[\xi_1, \xi_2],
\end{equation}
where the central term $K[\xi_1, \xi_2]$ depends on the dynamical
fields only through their boundary values.

The four dimensional low-energy Lagrangian obtained from string
theory is
\begin{equation}\label{L2}
{\mathbf{L}}_{abcd}=\epsilon_{abcd}\left[R-2(\nabla
\phi)^2-e^{-2\alpha\phi}F^{2}\right],
\end{equation}
where $\epsilon_{abcd}$ is the volume element, $\phi$ is the
dilaton scalar field, $F_{ab}$ is the Maxwell field associated
with a $U(1)$ sub-group of $E_8\times E_8$ or $Spin(32)/Z_2$, and
$\alpha$ is a free parameter which governs the strength of the
coupling of the dilaton to the Maxwell field. The reason we set
the remaining gauge fields and antisymmetric tensor field
$H_{\mu\nu\rho}$ to zero is that the metrics of stationary and
static dilaton black holes are almost obtained form the Lagrangian
(\ref{L2}). We know from Lagrangian (\ref{L2}) that the equations
of motion ${\mathbf{E}}$ for dynamical fields $A_\mu$, $\phi$, and
$g_{\mu\nu}$ can be respectively given by
 \be
 \nabla_\mu(e^{-2\alpha\phi}F^{\mu\nu})=0,\label{mf}\\
 \ee
 \be
 \nabla^2\phi+\frac{1}{2}e^{-2\alpha\phi}F_{\mu\nu}F^{\mu\nu}=0,
 \label{mp}\\
 \ee
 \be
 R_{\mu\nu}-\frac{1}{2}g_{\mu\nu}R=2\nabla_\mu\phi\nabla_\nu \phi
 -g_{\mu\nu}(\nabla\phi)^2+2e^{-2\alpha\phi}F_{\beta\nu}F^\beta_{\
 \mu}-\frac{1}{2}g_{\mu\nu}e^{-2\alpha\phi}F_{\mu\nu} F^{\mu\nu}.
 \label{Einstein}
 \ee
The symplectic potential (n-1)-form is
\begin{equation}\label{syp4}
{\mathbf{\Theta}}_{bcd}[g,
{\mathcal{L}}_{\xi}g]=4\epsilon_{abcd}\left\{\frac{1}{2}(\nabla _e
\nabla ^{[e}\xi ^{a]}+R_e^a \xi^e) -\xi^e\nabla_e\phi \nabla^a\phi
-e^{-2\alpha\phi}F^{a f}\left[F_{e f}\xi^e+(\xi^e
A_e)_{;f}\right]\right\}.
\end{equation}
From Eqs. (\ref{j1}) and (\ref{syp4}) we have
\begin{eqnarray}\label{j2}
{\mathbf{J}}_{bcd}&=&2\epsilon_{abcd}\left\{\nabla _e \nabla
^{[e}\xi ^{a]}-2e^{-2\alpha\phi} F^{a f}(\xi^e A_e
)_{;f}+\left[R_e^a-\frac{1}{2}\delta^a_eR -2\nabla_e \phi \nabla^a
\phi+\delta_e^a(\nabla\phi)^2 \right. \right. \nonumber
\\ & & \left. \left. -2e^{-2\alpha\phi}F^{a f}F_{e f}+\frac{1}{2}
\delta ^a_{e} e^{-2\alpha\phi}F^{2}\right]\xi^e\right\} \nonumber
\\&=&2\epsilon_{abcd}\left[\nabla _e \nabla ^{[e}\xi
^{a]}-2e^{-2\alpha\phi}F^{a f}(\xi^e A_e)_{;f}\right]\nonumber
\\ &=& 2\epsilon_{abcd}\left[\nabla _e \nabla ^{[e}\xi ^{a]}+4
\nabla_f (e^{-2\alpha\phi}\nabla^{[f}A^{a]}A_e \xi^ e)\right],
\label{j3}
\end{eqnarray}
in the second and third lines, we used the equations of motion
(\ref{Einstein}) and (\ref{mf}). Eqs. (\ref{dQ}) and (\ref{j3})
show that
\begin{equation}\label{Q1}
{\mathbf{Q}}_{cd}=-\epsilon_{abcd}\left[\nabla^{a}\xi^{b}+
4e^{-2\alpha\phi}A_e\xi^e\nabla^a A^b\right].
\end{equation}

For a stationary dilaton black hole, the dilaton scalar field, the
electromagnetic potential $ A_a $, and the Killing vector can be
respectively expressed as
\begin{eqnarray}
\phi&=&\phi(r, \theta),  \label{phi} \\ A_a&=&\Big(A_0(r, \theta),
\ A_1(r, \theta), \ A_2(r, \theta), \ A_3(r, \theta)\Big),
\label{Af}
\\
 \chi^a_H&=&\chi^{(t)}_H+\chi^{(\varphi)}_H=(1, \ 0,
\ 0, \ \Omega_H), \label{kl}
\end{eqnarray}
where the vector $\chi^{(t)}_H$ correspond to time translation
invariance, $\chi^{(\varphi)}_H$ to rotational symmetry, and
$\Omega_H=-(g_{t\varphi}/g_{\varphi\varphi})_H$ is the angular
velocity of the black hole.

As Carlip did in Ref. \cite{Carlip99} we define a ``stretched
horizon" $  \chi ^2=\epsilon$, where $\chi ^2=g_{ab}\chi^a\chi^b$,
$\chi^a$ is a Killing vector. The result of the computation will
be evaluated at the event horizon of the black hole by taking
$\epsilon$ to zero. Near the stretched horizon, one can introduce
a vector orthogonal to the orbit of $\chi^a$ by
$\nabla_a\chi^2=-2\kappa\rho_a, $ where $\kappa$ is the surface
gravity. The vector $\rho^a$ satisfies conditions
\begin{eqnarray}
 \chi^a\rho_a=-\frac{1}{\kappa}\chi^a\chi^b\nabla_a\chi_b=0,
 \qquad && \hbox{everywhere}\nonumber \\
 \rho^a\rightarrow \chi^a,\qquad && \hbox{at the horizon}.
\end{eqnarray}
To preserve ``asymptotic" structure at horizon, we impose Carlip's
boundary conditions\cite{Carlip99}
\begin{equation}\label{cond}
\delta \chi ^2=0,  \ \ \ \ \chi ^a t^b \delta g_{ab}=0, \ \ \ \
\delta \rho _a=-\frac{1}{2\kappa}\nabla _a(\delta \chi ^2)=0, \ \
\ \  at\ \ \ \chi^2=0,
\end{equation}
where $t^a$ is a any unit spacelike vector tangent to boundary
${\partial \mathbf{M}}$ of the spacetime ${\mathbf{M}}$.  And the
infinitesimal generator of a diffeomorphism is taken as
\begin{equation}\label{xi}
\xi^a ={\mathcal{R}}\rho^a+{\mathcal{T}}\chi ^a,
\end{equation}
where functions ${\mathcal{R}}$ and ${\mathcal{T}}$ obey the
relations \cite{Carlip99}
\begin{eqnarray}\label{rt}
{\mathcal{R}}=\frac{1}{\kappa}\frac{\chi^2}{\rho^2}\chi^a\nabla_a
{\mathcal{T}},\qquad && \hbox{everywhere} \nonumber \\
\rho^a\nabla_a{\mathcal{T}}=0, \qquad && \hbox{at the horizon}.
\end{eqnarray}
For a one-parameter group of diffeomorphism such that
$D{\mathcal{T}}_\alpha=\lambda _\alpha{\mathcal{T}}_ \alpha$, (
$D\equiv \chi^a\partial_a$ ), one introduces an orthogonality
relation\cite{Carlip99}
\begin{equation}\label{cond2}
\int_{\partial C}\hat{\epsilon}\
{\mathcal{T}}_\alpha{\mathcal{T}}_\beta \sim
\delta_{\alpha+\beta}.
\end{equation}
The technical role of the condition (\ref{cond2}) is to guarantee
the existence of generators $H[\xi]$. By using the other
future-directed null normal vector
$
N^a=k^a-\alpha \chi^a-t^a, $  with
$k^a=-\frac{1}{\chi^2}\left(\chi^a-\frac{|\chi|}{\rho}
\rho^a\right)$ and a normalization $N_a\chi^a=-1$, the volume
element can be expressed as
\begin{equation}\label{vol}
\epsilon_{abcd}=\hat{\epsilon}_{cd}(\chi_a N_b-\chi_b
N_a)+\cdots\cdots,
\end{equation}
the omitted terms do not contribute to the integral.

Form the right hand of  Eq. (\ref{bq})
 \begin{equation}\label{syp4a}
\int _{\partial C } \xi^b{\mathbf{\Theta}}_{bcd}=4\int _{\partial
C }\epsilon_{abcd}\xi^a \left\{\frac{1}{2}(\nabla _e \nabla
^{[e}\xi ^{b]}+R_e^b \xi^e)
-\xi^e\nabla_e\phi\nabla^b-e^{-2\alpha\phi}F^{f b}\left[F_{e
f}\xi^e+(\xi^e A_e)_{;f}\right]\right\},
\end{equation}
we know that the first two terms in the right hand of Eq.
(\ref{syp4a}) can be treated as Carlip did in Ref.
\cite{Carlip99}. At the horizon, by using Eqs. (\ref{phi}),
(\ref{Af}), (\ref{kl}) and (\ref{cond})- (\ref{vol}) we obtain
\begin{eqnarray}\label{phi21} & &\int _{\partial
C}\epsilon_{abcd} \xi^a_2\xi^e_1\nabla^b\phi\nabla_e\phi=0,
\end{eqnarray}
and
\begin{eqnarray} &
&\int _{\partial C}\epsilon _{abcd}e^{-2\alpha\phi}\xi^aF^{b
f}\left[F_{e f}\xi^e+(\xi^e A_e)_{;f}\right]\nonumber \\ &=&\int
_{\partial C}\epsilon _{abcd}e^{-2\alpha\phi}\xi^aF^{b
f}\delta_{\xi}A_f\nonumber
\\ &=&\int _{\partial C}\hat{\epsilon}
_{cd}e^{-2\alpha\phi}\left[\frac{|\chi|}{\rho}{\mathcal{T}}\rho_b+
\left(\frac{\rho}{|\chi|}+t\cdot\rho\right){\mathcal{R}}\chi_b\right]F^{b
f }\delta_{\xi}A_f\nonumber \\ &=&0.  \label{second}
\end{eqnarray}
Therefore we know that the last three terms in Eq. (\ref{syp4a})
also gives no contribution to $K[\xi_1, \xi_2]$.

By applying Eqs. (\ref{Af}), (\ref{kl}), (\ref{xi}), and
(\ref{vol}), we can show that, at the horizon, $ \int _{\partial C
}\epsilon_{abcd} e^{-2\alpha\phi}A_e\xi^e \nabla^a A^b \rightarrow
0.$  Hence, from Eq. (\ref{Q1}) we find
\begin{equation}\label{QQ}
\int _{\partial C }Q_{cd}=-\int _{\partial C
}\epsilon_{abcd}\nabla^a\xi^b.
\end{equation}

Denoting by $\delta _{\xi}$ the variation corresponding to
diffeomorphism generated by $\xi$, for the Noether current
 we have
$
\delta_{\xi_2}{\mathbf{J}}[\xi_1]= d[\xi_2
({\mathbf{\Theta}}[\phi, {\mathcal{L}}_{\xi_{1}}
\phi]-\xi_{1}\cdot {\mathbf{L}})].
$
Substituting it into Eq. (\ref{dh}) and using Eq. (\ref{syp4}) we
obtain
\begin{eqnarray}\label{dh1}
 \delta _{\xi_2}H[\xi_1] &=&\int_{\partial
 C}\left(\xi_2{\mathbf{\Theta}}[\phi, {\mathcal{L}}_{\xi_1}\phi
 ]-\xi_1{\mathbf{\Theta}}[\phi, {\mathcal{L}}_{\xi_2}\phi
 ]-\xi_2\xi_1{\mathbf{L}}\right) \nonumber
 \\&=&
 \int _{\partial C}\epsilon_{abcd}\left[
 \xi^a_2\nabla_e(\nabla^e\xi^b_1-\nabla^b\xi^e_1)-
 \xi^a_1\nabla_e(\nabla^e\xi^b_2-\nabla^b\xi^e_2)\right]\nonumber
 \\
 & &-  4\int _{\partial C}\epsilon_{abcd}e^{-2\alpha\phi}
 \left\{\xi_2^aF^{f b}\left[F_{e f}\xi_1^e+(\xi_1^e
 A_e)_{;f}\right]-\xi_1^aF^{f b}\left[F_{e f}\xi_2^e+(\xi_2^e
 A_e)_{;f}\right] \right\}\nonumber
 \\ & &-\int _{\partial C}\epsilon_{abcd}
 \left[4R^b_e(\xi^a_1\xi^e_2-\xi^a_2\xi^e_1)+\xi^a_2\xi^b_1
 {\mathbf{L}}\right]\nonumber \\
 & &-4\int _{\partial C}\epsilon_{abcd}\left(\xi^a_2\xi^e_1
  -\xi^a_1\xi^e_2\right)\nabla^b\phi\nabla_e\phi.
\end{eqnarray}
At the horizon, applying Eqs. (\ref{phi}), (\ref{Af}), (\ref{kl})
and (\ref{cond})- (\ref{vol}) we see that
\begin{eqnarray}\label{phi2} & &\int _{\partial
C}\epsilon_{abcd}
(\xi^a_2\xi^e_1-\xi^a_1\xi^e_2)\nabla^b\phi\nabla_e\phi\nonumber
\\ &=&\int _{\partial C}\hat{\epsilon}
_{cd}\left(\frac{1}{\kappa}\frac{\chi^2}{\rho^2}
\right)\left[\frac{|\chi|}{\rho}\rho_b\rho^e-\left(
\frac{\rho}{|\chi|}+t\cdot\rho\right)\chi_b\chi^e\right]
({\mathcal{T}}_2 D{\mathcal{T}}_1-{\mathcal{T}}_1
D{\mathcal{T}}_2)\nabla^b\phi\nabla_e\phi \nonumber
\\ &=&0,
\end{eqnarray}
\begin{eqnarray} & &\int _{\partial C}\epsilon_{abcd}
\xi^a_2\xi^b_1 {\mathbf{L}} \nonumber \\&=&\int _{\partial
C}\hat{\epsilon} _{c d
}{\mathbf{L}}\left[\frac{|\chi|}{\rho}{\mathcal{T}}_2\rho_b+
\left(\frac{\rho}{|\chi|}+t\cdot\rho\right){\mathcal{R}}_2\chi_b\right]
({\mathcal{T}}_1\chi^b+{\mathcal{R}}_1\rho^b)\nonumber \\ &=&\int
_{\partial C}\hat{\epsilon} _{c d}{\mathbf{L}} \left[
\frac{|\chi|}{\rho} {\mathcal{T}}_2{\mathcal{R}}_1\rho^2+
\left(\frac{\rho}{|\chi|}+
t\cdot\rho\right){\mathcal{R}}_2{\mathcal{T}}_1\chi^2\right]
\nonumber
\\&=&0,\label{lll}
\end{eqnarray}
and
\begin{eqnarray}\label{rrr}
& &\int _{\partial C}\epsilon_{abcd}
R^b_e(\xi^a_1\xi^e_2-\xi^a_2\xi^e_1)\nonumber \\ &=&\int
_{\partial C}\hat{\epsilon}
_{cd}R^b_e\left(\frac{1}{\kappa}\frac{\chi^2}{\rho^2}
\right)\left[\frac{|\chi|}{\rho}\rho_b\rho^e-\left(
\frac{\rho}{|\chi|}+t\cdot\rho\right)\chi_b\chi^e\right]
({\mathcal{T}}_1D{\mathcal{T}}_2-{\mathcal{T}}_2D
{\mathcal{T}}_1)\nonumber
\\ &=&0.
\end{eqnarray}
Substituting Eqs. (\ref{phi2}), (\ref{second}), (\ref{lll}) and
(\ref{rrr}) into Eq. (\ref{dh1}) we find
\begin{eqnarray}\label{dh1a}
\delta _{\xi_2}H[\xi_1]&=& \int _{\partial C}\epsilon_{abcd}\left[
\xi^a_2\nabla_e(\nabla^e\xi^b_1-\nabla^b\xi^e_1)-
\xi^a_1\nabla_e(\nabla^e\xi^b_2-\nabla^b\xi^e_2)\right].
\end{eqnarray}
We can interpret the left side of Eq. (\ref{dh1}) the variation of
the boundary term $J$ since the ``bulk" part of the generator
$H[\xi_1]$ on the left side vanishes on shell. On the other hand,
the change in $J[\xi_1]$ under a surface deformation generated by
$J[\xi_2]$ can be precisely described by Dirac bracket
$\{J[\xi_1], j[\xi_2]\}^*$ \cite{Carlip99}. Thus we have
\begin{equation}\label{dj3}
\{J[\xi_1], J[\xi_2]\}^*= \int _{\partial C}\epsilon_{abcd}\left[
\xi^a_2\nabla_e(\nabla^e\xi^b_1-\nabla^b\xi^e_1)-
\xi^a_1\nabla_e(\nabla^e\xi^b_2-\nabla^b\xi^e_2)\right].
\end{equation}
Inserting Eqs. (\ref{xi}), (\ref{rt}) and (\ref{vol}) into
(\ref{dj3}) we obtain
\begin{eqnarray}\label{dj5}
\{J[\xi_1], J[\xi_2]\}^*&=& -\int_{\partial
C}\hat{\epsilon}_{cd}\left[\frac{1}{\kappa}({\mathcal{T}}_1
D^3{\mathcal{T}}_2 -{\mathcal{T}}_2D^3{\mathcal{T}}_1)-2\kappa
({\mathcal{T}}_1D{\mathcal{T}}_2
-{\mathcal{T}}_2D{\mathcal{T}}_1)\right].
\end{eqnarray}
For any one-parameter group of diffeomorphism satisfying
conditions (\ref{xi}) and (\ref{rt}), it is also easy to check
that
\begin{equation}\label{xixi}
\{\xi_1,\xi_2\}^a=({\mathcal{T}}_1D{\mathcal{T}}_2
-{\mathcal{T}}_2D{\mathcal{T}}_1)\chi^a+\frac{1}
{\kappa}\frac{\chi^2}{\rho^2} D({\mathcal{T}}_1D{\mathcal{T}}_2
-{\mathcal{T}}_2D{\mathcal{T}}_1)\rho^a.
\end{equation}

The Hamiltonian (\ref{H}) consists of two terms, but Eqs
(\ref{phi2}) and (\ref{second}) and discussion about
$\xi\cdot{\mathbf{\Theta}}$ in Ref. \cite{Carlip99} show that the
second terms make no contribution. Then, we have
\begin{equation}\label{j4}
J[\{\xi_1, \xi_2\}]=\int_{\partial
C}\hat{\epsilon}_{cd}\left[2\kappa({\mathcal{T}}_1D{\mathcal{T}}_2
-{\mathcal{T}}_2D{\mathcal{T}}_1)-\frac{1}{\kappa}
D({\mathcal{T}}_1D^2{\mathcal{T}}_2
-{\mathcal{T}}_2D^2{\mathcal{T}}_1)\right].
\end{equation}
On shell  Eq. (\ref{algeb}) can be expressed as
\begin{equation}\label{algeb1}
\{J[\xi_1], J[\xi_2]\}^*=J[\{\xi_1, \xi_2\}]+K[\xi_1, \xi_2].
\end{equation}
Therefore, we know that from Eqs. (\ref{dj5}) and  (\ref{j4}) the
central term is
\begin{equation}\label{kk}
K[\xi_1, \xi_2]=\int_{\partial
C}\hat{\epsilon}_{cd}\frac{1}{\kappa}({D
\mathcal{T}}_1D^2{\mathcal{T}}_2-
D{\mathcal{T}}_2D^2{\mathcal{T}}_1).
\end{equation}
It is interesting to note that the constraint algebra
(\ref{algeb1}) with Eqs. (\ref{dj5}), (\ref{j4}), and (\ref{kk})
has same form as that for the vacuum case\cite{Carlip99}. In next
section, we will study statistical-mechanical entropies of some
stationary dilaton black holes by using the constraint algebra and
conformal field theory methods.

\vspace*{0.0cm}
\section{Statistical Entropy of stationary dilaton black hole}

In order to construct a standard Virasoro subalgebra from
constraint algebra (\ref{dj5}) and (\ref{j4})-(\ref{kk}), as
Cadoni, Mignemi and Carlip did in references \cite{Cadoni}
\cite{Carlip99} we define a new generator $\int d v J $ in which
the function $v$ takes period $T$.   Form stationary conditions
(\ref{kl}) we know that a one-parameter group of diffeomorphism
satisfying Eqs. (\ref{cond2}) and (\ref{xixi}) can be taken as
\begin{equation}\label{btk}
{\mathcal{T}}_n=\frac{T}{2\pi}exp\left[in(\frac{2\pi}{T} v +
C_\alpha (\varphi-\Omega_H v) )\right],
\end{equation}
where $C_\alpha$ is a arbitrary constant. We should note that
one-parameter group (\ref{btk}) is also valid for static black
hole since it is a special case of the stationary black hole with
$\Omega_H=0$. Substituting Eq. (\ref{btk}) into central term
(\ref{kk}) and using condition (\ref{cond2}) we obtain
\begin{equation}\label{bkkk}
  K[{\mathcal{T}}_m, {\mathcal{T}}_n]=-\frac{iA_H}{8\pi
  }\frac{2\pi}{\kappa  T}m^3\delta_{m+n, 0},
\end{equation}
where $A_H=\int _{\partial C}{\widehat{\epsilon}_{c d}}$ is the
area of the event horizon. Eq. (\ref{algeb1}) thus takes standard
form of a Virasoro algebra
\begin{equation}\label{algeb2k}
i\{J[{\mathcal{T}}_m],
J[{\mathcal{T}}_n]\}=(m-n)J[{\mathcal{T}}_{m+n}]+
\frac{c}{12}m^3\delta_{m+n, 0},
\end{equation}
with central charge
 \be\label{cccc}
\frac{c}{12}=\frac{A_H}{8\pi}\frac{2\pi}{\kappa T}.
 \ee
The boundary term $J[{\mathcal{T}}_0]$ can easily be  obtained by
using Eqs (\ref{dQ}), (\ref{Q1}), and (\ref{btk}), which is given
by
 \be\label{ddd}
J[{\mathcal{T}}_0]=\triangle=\frac{A_H}{8\pi}\frac{\kappa
T}{2\pi}.
 \ee
From standard Cardy's formula \cite{Carlip99}
\begin{equation}\label{brhok}
\rho(\triangle)\sim exp\left\{2\pi
\sqrt{\frac{c}{6}\left(\triangle-\frac{c}{24}\right)} \right\},
\end{equation}
we know that the number of states with a given eigenvalue
$\triangle$ of $J[{\mathcal{T}}_0]$ grows asymptotically for large
$\triangle$ as
\begin{equation}\label{brhok1}
\rho(\triangle)\sim exp\left[\frac{A_H}{4}\sqrt{2-\left(
\frac{2\pi}{\kappa T}\right)^2}\right].
\end{equation}
Only if we take the period $ T$ as the periodicity of the
Euclidean black hole, i.e., \be \label{bbb}
 T=\frac{2\pi}{\kappa}, \ee
the statistical entropy of the stationary dilaton black hole
\begin{equation}\label{bsk}
S_0\sim ln \rho(\triangle)=\frac{A_H}{4},
\end{equation}
coincides with the standard Bekenstein-Hawking entropy.

\vspace*{0.5cm}
\section{Logarithmic corrections to black hole entropy}\label{ncd}

\vspace*{0.5cm} Now lets us consider the first-order quantum
correction to the entropy. In oder to do that, we should first
derive the logarithmic corrections to the Cardy formula.

In references \cite{Carlip98a}\cite{Carlip00}, Carlip showed that
the number of states is
 \be \label{rho}
 \rho(\triangle)=\int d \tau e^{-2\pi i \triangle \tau}
 e^{-2\pi i \triangle_0
 \frac{1}{\tau}} e^{\frac{2 \pi i c}{24}\tau}
 e^{\frac{2 \pi i c}{24}\frac{1}{\tau}}\tilde{Z}(-1/\tau),
 \ee
where $\tilde{Z} (-1/\tau)$ approaches to a constants,
$\rho(\triangle_0)$, for large $\tau$. So the integral (\ref{rho})
can be evaluated by steepest descent provided that the imaginary
part of $\tau$ is large at the saddle point.

The integral takes the form
 \be \label{ii}
 I[a, b]=\int d \tau e^{2 \pi i a \tau+\frac{2\pi i
 b}{\tau}}f(\tau).
 \ee
The argument of the exponent is extremal at
$\tau_0=\sqrt{\frac{b}{a}}$, and expanding around $\tau_0$, one
has \cite{Carlip00}
 \be\label{iii}
  I[a, b]\approx \int d \tau e^{4 \pi i a \sqrt{ab}+\frac{2\pi i
 b}{\tau_0^3}(\tau-\tau_0)^2}f(\tau_0)=\left(-\frac{b}{4a^3}
 \right)^{1/4}e^{4\pi i \sqrt{ab}}.
 \ee
Comparing Eqs. (\ref{rho}) with (\ref{ii}) we know
 \be
 a=\frac{c}{24}-\triangle, \ \ \ b=\frac{c}{24}-\triangle_0.
 \ee
Therefore, for large $ \triangle$, if we let $
c_{eff}=c-24\triangle_0, $ the number of states can be expressed
as
 \be
\label{ccc} \rho_{cq}(\triangle)\approx \left[\frac{c_{eff}}{96
\left(\triangle-\frac{c}{24}\right)^3}\right]^{1/4} exp\left\{2\pi
\sqrt{\frac{c_{eff}}{6}\left(\triangle-\frac{c}{24}\right)}
\right\}\rho(\triangle_0).
 \ee
The exponential part in (\ref{ccc}) gives the  Carlip's result
(C.3) in Appendix C in Ref. \cite{Carlip99}, the factor before the
exponent devotes the logarithmic  correction to black hole
entropy.

By Using the central charge (\ref{cccc}), eigenvalue (\ref{ddd}),
constraint condition of the period (\ref{bbb}), and new Cardy
formula (\ref{ccc}), we know that the statistical entropy
including first-order quantum correction is given by
 \ba
 S&=&\frac{A_H}{4}-\frac{3}{2}\ln \frac{A_H}{4}+\ln
 c+const.,\nonumber \\
&=&\frac{A_H}{4}-\frac{1}{2}\ln \frac{A_H}{4} +const.\  .
  \ea
The first line has two logarithmic terms and  agrees with Carlip's
results (\ref{Centropy}) \cite{Carlip00}. However, after we take $
T=\frac{2\pi}{\kappa}$, the second shows that the factor of the
logarithmic term becomes $-\frac{1}{2}$, which is different from
Kaul and Majumdar's result $-\frac{3}{2}$.

\vspace*{0.0cm}

\section{summary and discussion}

We extend Carlip's investigation in Ref. \cite{Carlip99}  to four
dimensional low-energy string theory by the covariant phase
techniques. With Carlip's boundary conditions, a standard Virasoro
subalgebra with corresponding central charge for stationary
dilaton black hole is constructed at a Killing horizon. We find
that only we take $ T$ as the periodicity of the Euclidean black
hole, $T=\frac{2\pi}{\kappa}$, the statistical entropy of the
stationary dilaton black hole yielded by standard Cardy formula
agrees with its Bekenstein-Hawking entropy. Therefore, Carlip's
conclusion---the asymptotic behavior of the density of states may
be determined by the algebra of diffeomorphism at horizon---is
valid for stationary dilaton black holes obtained from the
low-energy effective field theory with Lagrangian (\ref{L2}).

When we consider first-order quantum correction the entropy
contains extra logarithmic terms which agrees with Carlip's
results (\ref{Centropy}) \cite{Carlip00}. However, from above
discussions we know that in order to get the Bekenstein-Hawking
entropy we have to take $ T=\frac{2\pi}{\kappa}$. That is to say,
we can not set central charge $c$ to be a universal constant,
independent of area of the event horizon, by adjusting periodicity
$ T$ as Carlip suggested in Ref. \cite{Carlip00}. Therefore, the
factor of the logarithmic term is $-\frac{1}{2}$, which is
different from Kaul and Majumdar's result, $-\frac{3}{2}$.

From the derivation given in the section \ref{ncd} we know that
the new Cardy formula (\ref{ccc}) is valid for general black hole
whether or not the black hole is dilatonic. Hence, the factor of
the logarithmic term will be $-\frac{1}{2}$ as long as the
spacetime is such that (\ref{kk}) is obey. This means that the
discrepancy between Carlip's \cite{Carlip00} approach and that of
Kaul and Majumdar \cite{Kaul00} is not just for the dilaton black
hole, but for any black hole which respects (\ref{kk}), where T is
the periodicity of the Euclidean black hole.

\vspace*{0.5cm}

\begin{acknowledgements}
This work was supported in part by the National Natural Science
Foundation of China under grant No. 19975018, Theoretical Physics
Foundation of China under grant No. 19947004 and National
Foundation of China through C. N. Yang.
\end{acknowledgements}

\end{document}